\documentclass[extra]{gji2}
\usepackage{amsmath, amssymb, graphicx}
\usepackage{hyperref}
\usepackage{lineno}

\DeclareMathOperator{\erfc}{erfc}

\title{Fluid pressure heterogeneity during fluid flow in rocks: New laboratory measurement device and method}
\author[N. Brantut and F. M. Aben]{Nicolas Brantut and Franciscus M. Aben\\Department of Earth Sciences\\University College London, London, UK}
\date{\ }

\begin{document}

\maketitle

\begin{summary}
  We present a new type of transducer capable of measuring local pore fluid pressure in jacketed rock samples under elevated confining pressure conditions. The transducers are passive (strain-gauge based), of small size (7~mm in diameter at the contact with the rock and around 10~mm in length), and have minimal dead volume (a few mm$^3$). The transducers measure the differential pressure between the confining fluid and the internal pore pressure. The design is easily adaptable to tune the sensitivity and working pressure range up to several hundred megapascals. An array of four such transducers was tested during hydrostatic pressurisation cycles on Darley Dale sandstone and Westerly granite. The prototypes show very good linearity up to 80~MPa with maximum deviations of the order of 0.25~MPa, regardless of the combination of pore and confining pressure. Multiple internal pore pressure measurements allow us to quantify the local decrease in permeability associated with faulting in Darley Dale sandstone, and also prove useful in tracking the development of pore pressure fronts during transient flow in low permeability Westerly granite.
\end{summary}

\begin{keywords}
  Permeability and porosity; Hydrogeophysics; Geomechanics
\end{keywords}

\section{Introduction}

Pore fluid pressure is a key physical variable exerting first order controls on properties such as strength, transport properties, and seismic wave speeds in rocks. Since the early 1960s, rock deformation experiments have commonly been conducted with concomitant measurements or control of pore pressure, typically using external servo-controlled pumps connected to the rock samples' pore space by some length of high pressure tubing \citep[][Chap. 2]{paterson05}. During laboratory experiments, the pore space of rock samples is compacted or dilated due to elastic and inelastic deformation. When a constant fluid pressure is imposed at the ends of a specimen, the pore pressure remains homogeneous throughout the specimen provided that the pore pressure diffusion rate in the sample is much larger than the rate of pore pressure change due to pore space dilation/compaction. This is the ``drained'' condition. As a rule of thumb, for deformation at a rate $\dot{\varepsilon}$ of a sample of size $L$ with hydraulic diffusivity $\alpha$, drained conditions are ensured when $\alpha/(L^2\dot{\varepsilon})\gg 1$. By contrast, when pore space deformation is too rapid compared to fluid flow rate, pore pressure is expected to vary within the rock, and the externally imposed pressure is no longer representative of the internal state of the sample. Such ``partially drained'' conditions are typically avoided in laboratory experiments since the pore pressure is then unknown and nonuniform within the specimen. However, many key deformation processes occur very rapidly compared to drainage timescales. This is the case, for instance, of rock failure and stick slip. Our understanding of such rapid processes would significantly improve by performing \emph{internal} pore pressure measurements during high pressure laboratory experiments. Indeed, recent laboratory experiments using specific sample arrangements have shown that local pressure measurements provide key information on the dynamics of failure and fault slip \citep{brantut20, proctor20} in partially drained conditions.

Separately from deformation-induced pore fluid flow, the characterisation of hydraulic transport properties of rocks requires pore pressure gradients to develop across rock samples. In the simplest case of constant flow rate in a homogeneous permeable rock, Darcy's law predicts that a linear pore pressure gradient develops. However, nonuniform gradients are expected if the permeability of the material varies spatially, or if permeability itself depends on pore pressure \citep[e.g.,][]{rice92b}. Nonlinear pore pressure profiles are also expected in homogeneous materials during characterisation with transient or oscillatory testing methods, such as the pulse-decay \citep[e.g.,][]{brace68,hsieh81,bourbie82} or the sinusoidal steady-state oscillation method \citep[e.g.,][]{kranz90,fischer92}. While those methods have been developed to work efficiently in experimental configurations with only upstream and downstream pore pressure measurements, the estimation of transport properties is complicated by the ``dead'' fluid volume in the up- or downstream tubing and reservoirs. Fluid reservoirs connected to the pore space of the rock artifically provide additional poro-elastic compliance, which can impact both steady-state and transient measurements \citep[e.g.,][]{bernabe06,pimienta16}. Internal pore pressure measurements at different locations along rock samples during transient flow could help reduce uncertainties associated with the presence of dead volumes, and provide direct evidence for local heterogeneities affecting fluid flow.

Internal pore pressure measurements are therefore desirable during high pressure laboratory rock deformation and characterisation experiments. For internal measurements to be achievable and useful, they must be conducted with passive transducers with minimal additional ``dead'' volume \citep[e.g.,][]{hart01}. Recent progress in this direction has been made with the development of fibre optic sensors \citep{reinsch12, blocher14, nicolas20}. While fibre optic pressure transducers have clear benefits due to their small size and lack of sensitivity to electromagnatic noise, their manufacture and interrogation requires specific equipment and technology that is not commonly available in rock physics laboratories. Here, we present in detail a design and test results of strain gauge-based pore pressure transducers that can be fitted on rock samples inside a conventional triaxial apparatus. Similar transducers were used by \citet{brantut20} to measure dilatancy-induced pressure changes in faults during rupture and slip. Here, we elaborate further on the transducer design, show its limitations and present two practical uses to (1) determine permeability heterogeneity in a faulted sandstone and (2) to estimate both permeability and storage capacity during transient pore pressure steps in thermally cracked Westerly granite.


\section{Transducer design}

\subsection{Concept and construction}
\label{sec:concept}

\begin{figure}
  \centering
  \includegraphics{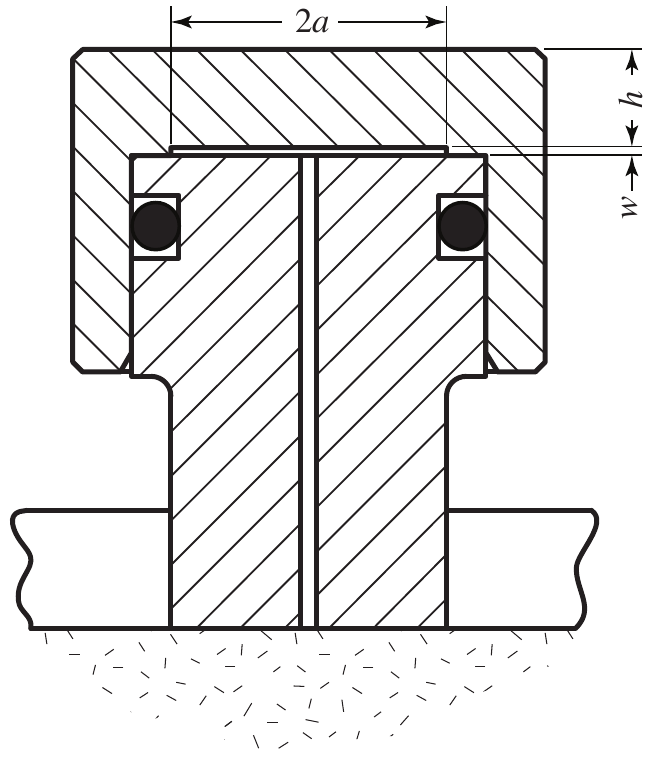}
  \caption{Drawing of the pressure transducer. The unsupported part of the cap has a thickness $h$ and radius $a$, and the open gap in the cavity has thickness $w$. Confining pressure $P_\mathrm{c}$ is applied outside the sealed cap, and the cavity is connected to the sample's pore space through a small radius conduit. Here, $h=2.5$~mm, $a=3.5$~mm, $w=0.2$~mm and the conduit has a radius of $0.2$~mm.}
  \label{fig:sensor}
\end{figure}

A schematic of the transducer is shown in Figure \ref{fig:sensor}. The transducer consists of two separate parts: (1) a stem with a small diameter conduit connecting a curved surface in contact with the rock and an open face at the top, with o-ring housing in the top part, and (2) a cap with a thin shoulder on the internal rim forming a penny-shaped cavity with the upper surface of the stem. The stem is mounted directly on the rock sample's surface, and sealed from the confining fluid with epoxy. An o-ring seals the upper part of the stem, making the penny-shaped cavity connected to the sample's pore space and isolated from the confining medium. The top surface of the cap is mounted with a diaphragm strain gauge recording the elastic distortion of the cap driven by pressure differentials between confining (outside) and pore (in the isolated cavity) fluid.

Using a thin elastic plate model, the radial and tangential elastic strains at the top of the unsupported part of the cap (i.e., above the penny-shaped cavity) can be approximated as:
\begin{linenomath}\begin{align}
  \varepsilon_{r} &= -\frac{3P_\mathrm{diff}(1-\nu^2)(a^2-3r^2)}{8Eh^2},\\
  \varepsilon_{\theta} &=-\frac{3P_\mathrm{diff}(1-\nu^2)(a^2-r^2)}{8Eh^2},
\end{align}
\end{linenomath}
respectively, where $P_\mathrm{diff} = P_\mathrm{c}-P_\mathrm{f}$ is the difference between confining $P_\mathrm{c}$ and internal fluid pressure $P_\mathrm{f}$, $\nu$ is the Poisson ratio of the cap material, $E$ is its Young's modulus, $h$ is the thickness of the cap, and $a$ is the radius of the unsupported surface. A diaphgram strain gauge with a circular pattern is bonded to the cap, and wired in a full bridge configuration to output the difference in the radial and tangential strains (averaged over each sensing element). Assuming a gauge factor of 2.0, the total gauge output is expressed as (see Tech Report from \citet{vishay10})
\begin{linenomath}\begin{equation} \label{eq:eout}
  e_\mathrm{out} \approx 0.75\frac{P_\mathrm{diff}a^2(1-\nu^2)}{Eh^2}\times 10^3\quad\mathrm{mV/V}.
\end{equation}\end{linenomath}
The maximum elastic deflection of the cap as its center is given by
\begin{linenomath}\begin{equation} \label{eq:deflection}
  d = \frac{3P_\mathrm{diff}a^4(1-\nu^2)}{16Eh^3}.
\end{equation}\end{linenomath}

We built our prototype transducers out of stainless steel PH-17-4, with elastic parameters $E=210$~GPa and $\nu=0.285$. For a design pressure of $P_\mathrm{diff}=100$~MPa and a cavity of radius $3.5$~mm, a thickness $h=2.5$~mm ensures that the maximum strain in the cap remains less than $10^{-3}$, so that the sensing element remains well within the elastic regime. For that geometry, the maximum deflection at the center of the cap is of the order of a few microns. We used diaphragm strain gauges from BCM, model ECF-350-7KA-B-(11)-O-SP, with gauge factor of 2.1. The predicted transducer sensitivity is of $e_\mathrm{out}\approx 0.64$~mV/V, using Equation \eqref{eq:eout}.

The dead volume in the cavity can be minimised by reducing the machined recess down to a fraction of a millimetre. Similarly, the conduit volume should also be minimised by producing a hole as small as technically possible. We constructed the transducer using a recess $w=0.2$~mm (machined with conventional cutting tools) and by laser-drilling the conduit at a diameter of $0.4$~mm. With an unsupported cavity radius of $3.5$~mm, the resulting dead volume in the whole transducer is $V_\mathrm{trans}=2.89$~mm$^3$.

\subsection{Flow disturbance due to dead volume}

While the dead volume inside the transducer is small, it is not zero. We therefore expect the transducer to introduce transient disturbances of the pore pressure field in the sample. The pore pressure recorded by the transducer (i.e., the fluid pressure inside the transducer's cavity) is therefore different from what the pressure would be in the absence of any transducer.

\begin{figure*}
  \centering
  \includegraphics{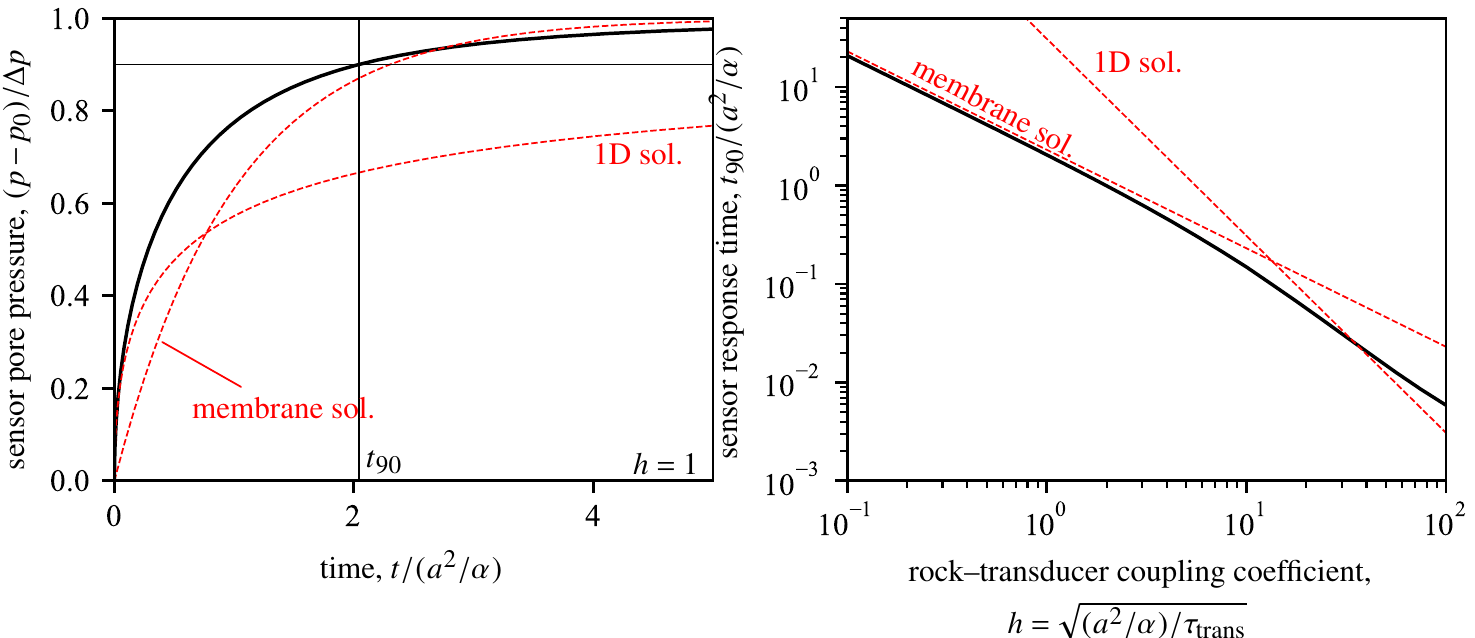}
  \caption{Left: Fluid pressure in the sensor cavity as a function of time, in response to a step change in pressure in the neighbouring rock. Rock-transducer coupling parameter $h=1$. Solid black line is a full 2D axisymmetric numerical solution, and two approximations are shown in red dashed lines (see Appendix \ref{ax:0}). Response time $t_{90}$ is defined by the time when the sensor pressure reaches 90\% of the pore pressure in the rock. Right: Sensor response time as a function of coupling parameter $h$. Solid black line shows numerical solution and approximations are shown in red dashed lines.
  \label{fig:filter}}
\end{figure*}

At the contact between the transducer stem and the sample, the fluid mass balance dictates the evolution of pore pressure following
\begin{linenomath}\begin{equation} \label{eq:BCdim}
  \frac{\partial p}{\partial t} - \frac{k A}{\eta\beta_\mathrm{trans}}\frac{\partial p}{\partial z} = 0,
\end{equation}\end{linenomath}
where $z$ is the coordinate perpendicular to the sample's surface, $k$ is the permeability of the sample, $A$ the area of contact between the sample and the transducer, $\eta$ is the fluid viscosity and $\beta_\mathrm{trans}$ is the storage capacity of the transducer. Assuming that the fluid is much more compressible than the transducer's cavity, $\beta_\mathrm{trans}$ can be approximated by $V_\mathrm{trans}\beta_\mathrm{f}$, where $\beta_\mathrm{f}$ is the fluid compressibility. This approximation yields a lower bound for $\beta_\mathrm{trans}$.

Choosing a timescale $\tau$ and normalising distances by $\sqrt{\alpha\tau}$, where $\alpha=k/(\eta\beta)$ is the sample's hydraulic diffusivity, the boundary condition \eqref{eq:BCdim} is rewritten as
\begin{linenomath}\begin{equation} \label{eq:BC}
  \frac{\partial p}{\partial (t/\tau)} - h \frac{\partial p}{\partial (z/\sqrt{\alpha \tau}) }=0 ,
\end{equation}\end{linenomath}
where 
\begin{linenomath}\begin{equation} \label{eq:h}
  h =  \frac{A \sqrt{\alpha \tau} \beta}{\beta_\mathrm{trans}}
\end{equation}\end{linenomath}
is the rock-transducer coupling coefficient, denoting $\beta$ the storage capacity of the sample.

\begin{table*}
  \small
  \centering
    \caption{Filtering characteristics of the transducer for a range of rocks. Data extracted from \citet[][Table 7.2]{jaeger07}. Other parameters are $A=38.5$~mm$^2$, $\eta=10^{-3}$~Pa~s and $\beta_\mathrm{trans}=V_\mathrm{trans}\beta_\mathrm{f}=1.45\times10^{-9}$~mm$^3$Pa$^{-1}$, where $\beta_\mathrm{f}=5\times10^{-10}$~Pa$^{-1}$ is the fluid compressibility.}
  \label{tab:filterrocks}
  \begin{tabular}{ccccccc}
    \hline
    Rock type & Permeability & Storage capacity & Disturbed volume & Critical time & Response time\\
              & $k$ & $\beta$ & $\beta_\mathrm{trans}/\beta$ & $\tau_\mathrm{trans}$   & $t_{90}$\\
              & (m$^2$) & ($\times10^{-9}$ Pa$^{-1}$) & (mm$^3$) &(s)  & (s)\\
    \hline
    Berea sandstone & $1.9\times10^{-13}$ & $2.1\times10^{-1}$ & $6.9$ & $6\times10^{-8}$  & $2\times10^{-6}$\\
    Boise sandstone & $8\times10^{-13}$ & $1.2\times10^{-1}$ & $12.1$ & $8\times10^{-9}$  & $2.8\times10^{-7}$\\
    Westerly granite & $4\times10^{-19}$ & $1.8\times10^{-2}$ & $80.6$ & $0.2$ & $0.8$\\
    Tennessee marble & $10^{-19}$ & $9.2\times10^{-3}$ & $157.6$ & $1.5$ & $3.0$\\
    \hline
  \end{tabular}
\end{table*}

Inspecting \eqref{eq:BC}, we expect the contact between sample and transducer to be approximately governed by a ``no flow'' condition if $h\gg1$, in which case the dead volume of the transducer has a negligible effect on the pore pressure field. Conversely, for $h\ll1$, the boundary condition is approximately that of constant pressure, so that the transducer dampens all pore pressure variations. Rewriting $h=\sqrt{\tau/\tau_\mathrm{trans}}$, where $\tau_\mathrm{trans}=\beta_\mathrm{trans}^2\eta/(A^2 k \beta)$, it becomes clear that the transducer dampens pore pressure variations occurring over timescales shorter than $\tau_\mathrm{trans}$, and that accurate measurements can only be achieved at timescales much greater than $\tau_\mathrm{trans}$. In terms of design, minimising $\tau_\mathrm{trans}$ to obtain a fast response time and minimal disturbance of the pressure field requires small sensor storage and large sensor area. Unfortunately, the critical timescale $\tau_\mathrm{trans}$ is not solely a function of the transducer dimensions, but also includes contributions from the sample storage capacity and permeability, so that the transducer response is not universal.

The disturbances are expected to involve small rock volumes in the vincinity of the transducer. An order of magnitude estimate for the rock volume where pore pressure is disturbed due to the presence of transducer is given by the ratio of storage capacities $\beta_\mathrm{trans}/\beta$. Table \ref{tab:filterrocks} gathers representative values of this disturbed volume for a range of rock types. For porous sandstones, only a few mm$^3$ of the rock would be impacted by the presence of the transducer, while the disturbed volume is higher, up to $100$~mm$^3$, in very tight marble or granite. For sample sizes of 4~cm in diameter and 10~cm in length, such disturbed volumes are negligible fractions ($\lesssim 0.1$\%) of the overall sample volume, so it is unlikely that overall flow patterns would be changed by the presence of transducers.

In order to obtain a clearer, quantitative view of the flow disturbances due the transducer beyond the basic dimensional considerations given above, we have to specialise our analysis. A number of realistic experimental scenarios can be envisioned: transient flow due to imposed remote pore pressure step or oscillations, distant pore pressure sources in a given volume and over a range of timescales, etc. Rigorous modelling of the effect of transducers on the pore pressure field would require a specific analysis for each case, and a generic treatment cannot be given. Here, we choose to analyse in full a simple but revealing ``elementary'' problem, that of the pore pressure disturbance associated with the presence of the transducer in response to a uniform, sudden change in pore pressure in the rock. This situation corresponds, for instance, to the case of a rapid porosity change due to inelastic dilatancy or poroelastic Skempton effect in response to a step change in stress. Of interest here is how quickly the pressure measured by the sensor equilibrates with the pressure within the rock.

\begin{figure*}
  \centering
  \includegraphics{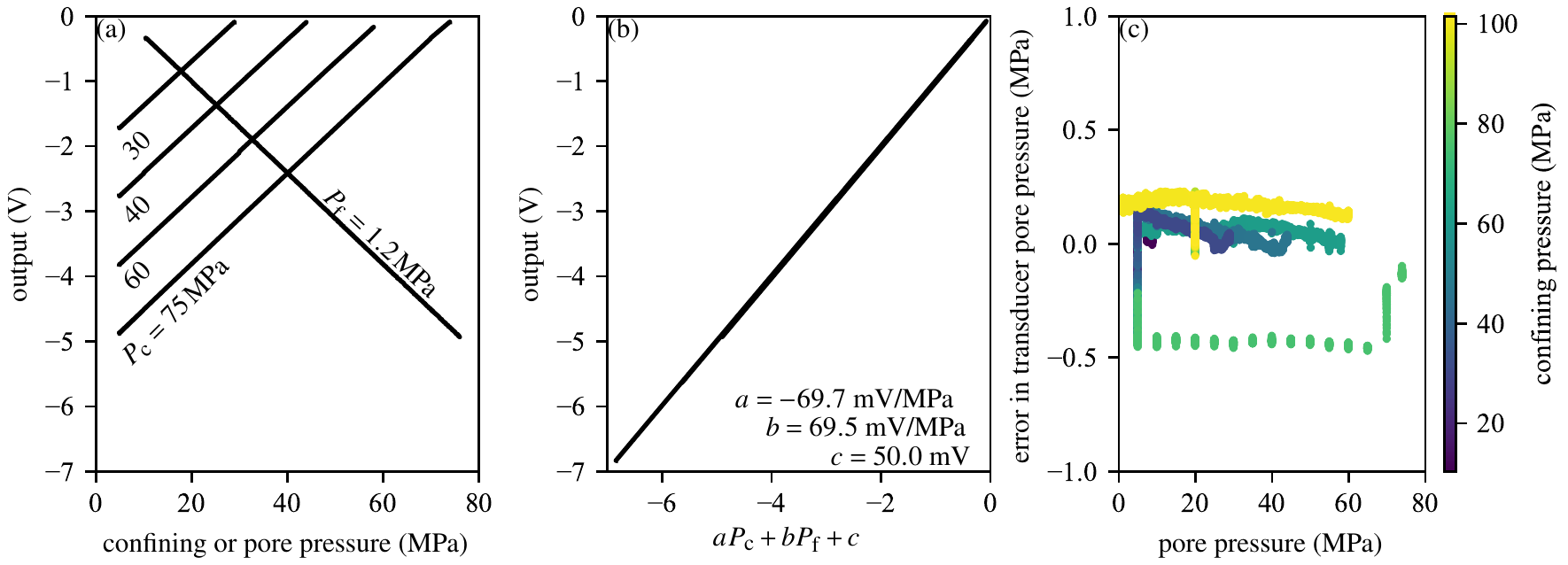}
  \caption{Calibration results for transducer FB1 mounted on Darley Dale sandstone. (a) Amplified output data for a range of $P_\mathrm{c}$ and $P_\mathrm{f}$. (b) Amplified output data fitted by the bilinear relationship \eqref{eq:V}. (c) Absolute error as a function of imposed pore pressure, for a range of confining pressure (indicated by color scale).}
  \label{fig:calibration}
\end{figure*}

\begin{table*}
  \centering
    \caption{Transducer calibration from least-square fitting to equation \eqref{eq:V}. Voltages are amplified output using input bridge voltage of 10~V and amplification ratio of $\times1000$.}
  \label{tab:coefs}
  \begin{tabular}{ccccc}
    \hline
    Transducer & $a$ (mV/MPa) & $b$ (mV/MPa) & $c$ (mV) & max. absolute error (MPa)\\
    \hline
    FB1 &  $-69.72$ & $69.5$  & $50.0$  & $0.47$\\
    FB2 &  $-68.9$  & $69.46$ & $37.0$  & $0.47$\\
    FB3 &  $-71.92$ & $68.25$ & $29.72$ & $0.56$\\
    FB4 &  $-70.27$ & $71.17$ & $71.95$ & $0.43$\\
    \hline
  \end{tabular}
\end{table*}

Let us consider that the rock occupies the half-space $z>0$, where the pore pressure is governed by
\begin{linenomath}\begin{equation} \label{eq:dp_2d}
  \frac{\partial p}{\partial t} - \alpha\left[\frac{\partial^2 p}{\partial z^2} + \frac{1}{r}\frac{\partial}{\partial r}\left(r\frac{\partial p}{\partial r}\right) \right],
\end{equation}\end{linenomath}
where $r$ is the radial position. The sensor is in contact at $z=0$ over the region $|r|<a$, where the boundary condition is given by \eqref{eq:BCdim}. The domain $|r|>a$ is governed by a zero flux condition in the $z$ direction to simulate the impermeable jacket at the surface of the sample. The pore pressure is assumed homogeneous, equal to $p_0$ at $t=0^-$, and at $t=0^+$ a sudden change $p=p_0+\Delta p$ is imposed in the whole region $z=0$. The natural timescale for the problem is $\tau=a^2/\alpha$, and we expect the evolution of pore pressure to depend on $h$, i.e., on the ratio of two timescales $\tau_\mathrm{trans}$ and $a^2/\alpha$. Equation \eqref{eq:dp_2d} is solved numerically for a range of $h$, and closed-form approximations are found at short and long timescales (see Appendix \ref{ax:0}, and Figure \ref{fig:filter}). At time small compared to $a^2/\alpha$, the sensor disturbance remains in narrow region near $z=0$, and the response is well modelled by a 1D approximation (Appendix \ref{ax:0}). At large time, the problem can be simplified by approximating the pore pressure gradient in \eqref{eq:BCdim} as $\partial p/\partial z \approx (p-p_0-\Delta p)/a$ (i.e., using a so-called ``membrane approximation'' of the gradient), which leads to
\begin{linenomath}\begin{equation} \label{eq:p_membrane}
  p(t)-p_0 = \Delta p \left(1 - e^{-t/\sqrt{\tau\tau_\mathrm{trans}}}\right),
\end{equation}\end{linenomath}
The sensor response time can be estimated by the time $t_{90}$ required for pore pressure to reach $90$\% of the uniformly applied step (Figure \ref{fig:filter}b). The membrane approximation \eqref{eq:p_membrane} is a very accurate representation of the sensor response over a wide range of parameter $h$, where $t_{90} = - \sqrt{\tau\tau_\mathrm{trans}}\times\log(0.1)$.

Table \ref{tab:filterrocks} gathers representative values of $\tau_\mathrm{trans}$ and $t_{90}$ for a range of rock types, using data summarised in \citet[][Table 7.2]{jaeger07}. Except for tight rocks with small storage capacity and permeability, the transducer is expected to have a negligible impact on pore pressure changes and should be able to capture fast variations.

\section{Experimental setup and calibration}


\subsection{Methods and materials}

Four prototype transducers were manufactured and tested in the triaxial Rock Physics Ensemble in the Rock and Ice Physics Laboratory at University College London.

Samples of two different rocks were used to test the transducers: Darley Dale sandstone (porosity of $13.3$\%, measured by the triple weight method) was chosen for its relatively high permeability and storage capacity \citep[e.g.,][]{zhu97}, and Westerly granite was chosen for its low permeability and storage capacity \citep[e.g.,][]{brace68b}. One sample of each rock was cored to produce cylinders of $40$~mm in diameter, and its ends ground parallel to a length of $100$~mm. The sample of Westerly granite was subsequently thermally treated at atmospheric pressure and $600^\circ$C to increase its nominal permeability \citep{nasseri09,wang13} and allow fluid saturation and drainage under reasonable laboratory timescales.



\begin{figure*}
  \centering
  \includegraphics{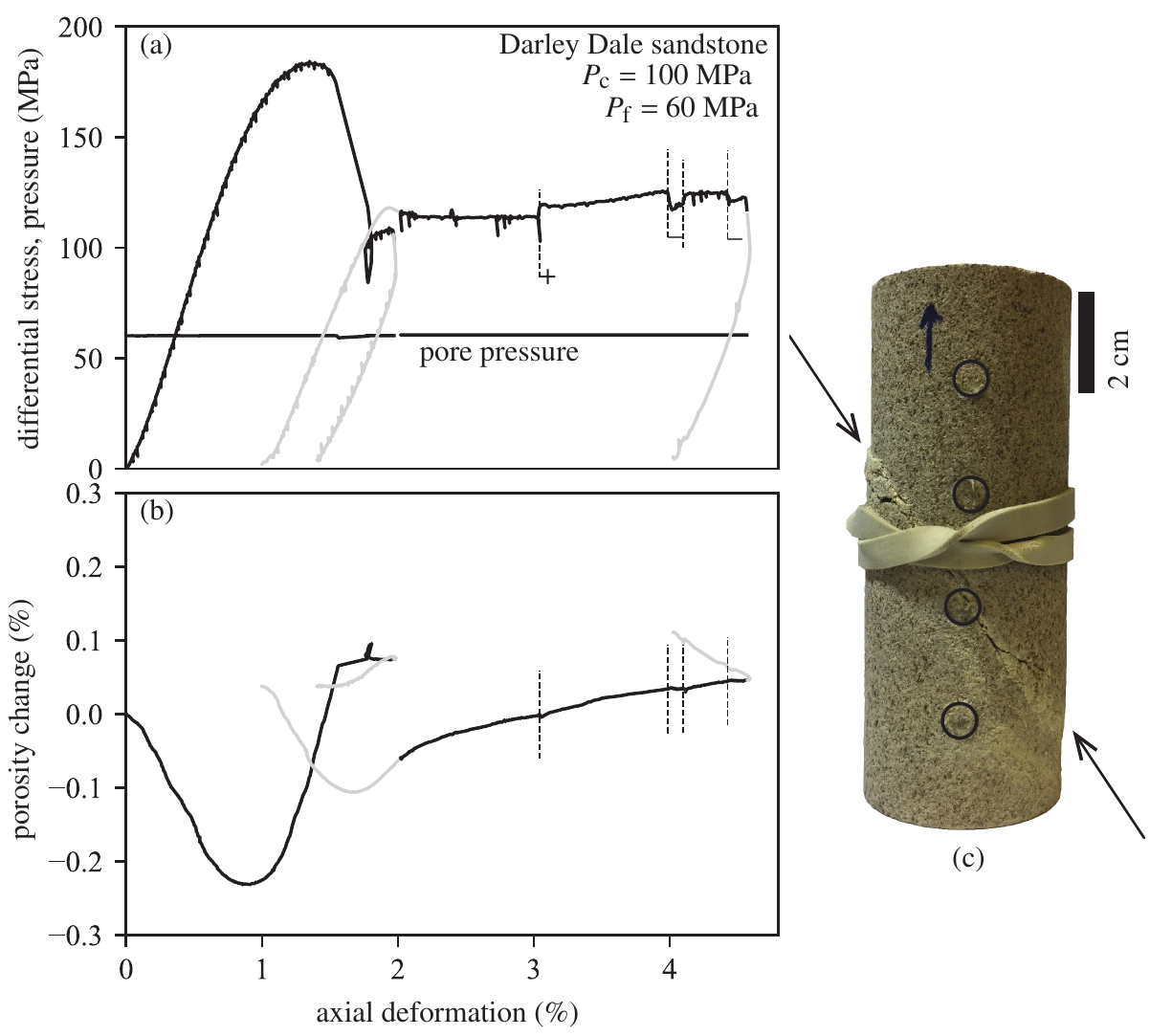}
  \caption{(a) Differential stress and internal pore pressure vs. axial deformation and (b) porosity change during rupture and slip. Internal pore pressure remained homogeneous in the sample throughout the test. Grey curves correspond to unloading and reloading phases. Dashed lines mark sequences where strain rate was increased ($+$) or decreases ($-$) by a factor of $10$. (c) Photograph of the deformed sample, showing the location of the shear fracture (diametric arrows) and the position of the pore pressure transducers (circles).}
  \label{fig:deformation}
\end{figure*}

The samples were jacketed in perforated $3$~mm thick nitrile sleeves. The four transducers were located every 20~mm along the samples' height. Jacket sealing was achieved by applying epoxy glue (Loctite EA 9455) at the interface between the jacket holes and transducer stems. 

The equipped samples were placed in the pressure vessel of the triaxial apparatus, and electrical connections were made by a set of high pressure leadthroughs. Each pressure transducer was wired to a broadband full bridge amplifier, with an input voltage of 10~V and amplification factor of $\times1000$. The amplified output was logged at $1$~Hz sampling rate, together with all other sensor outputs from the triaxial apparatus.


The pressure vessel was filled with silicone oil and the confining pressure was raised with an electric pump and intensifier, controlled within $0.4$~MPa, and recorded by a pressure transducer at the inlet of the pressure vessel (precision $0.01$~MPa). Pore pressure was imposed at either end of the sample using a servo-hydraulic intensifier equipped with a volumometer. The samples were initially dry, and were saturated in situ by flushing distilled water from the upstream end to the vented downstream end. Full saturation was achieved when steady-state flow was established (as measured from the intensifier volume) while water was collected on the vented side of the sample. The pore pressure was measured at both ends of the sample using transducers with $0.01$~MPa precision; transducer offsets and absolute accuracy were calibrated within $0.02$~MPa by ensuring $0.1$~MPa pressure reading at ambient pressure (vented pore circuit).

Axial deformation was imposed with a servo-hydraulic ram and piston. Load was measured with an external load cell and corrected for piston seal friction. Axial shortening was measured with a pair of linear variable differential transducers, corrected for elastic deformation of the loading column.

All tests were performed using nominally drained conditions (pore pressure maintained constant at both ends of the sample), except when explicitly mentioned otherwise.

\subsection{Calibration results}

The Darley Dale sandstone sample was pressurised using a range of confining pressures and pore pressures, keeping $P_\mathrm{f}<P_\mathrm{c}$ at all times. Variations in $(P_\mathrm{c},P_\mathrm{f})$ were sufficiently slow compared to the diffusion timescale so that a uniform pore pressure was ensured inside the sample. The amplified transducer readings showed a good linearity with both $P_\mathrm{c}$ and $P_\mathrm{f}$ (see example in Figure \ref{fig:calibration}a). The output $V$ in Volts was fitted to a linear relation (Figure \ref{fig:calibration}b)
\begin{linenomath}\begin{equation}\label{eq:V}
  V = aP_\mathrm{c} + bP_\mathrm{f} + c,
\end{equation}\end{linenomath}  
and best-fitting coefficients (in the least-square sense) are reported in Table \ref{tab:coefs}. The sensitivity to confining pressure is nearly equal in magnitude and opposite in sign to the sensitivity to pore pressure, which is expected from the construction of the transducer. In addition, the measured sensitivities are all around $0.7$~mV/V (in terms of raw output), which is close to the theoretical prediction made in Section \ref{sec:concept}. Using the linear relation \eqref{eq:V}, one can compute the local pore pressure in the sample at each sensor position as
\begin{linenomath}\begin{equation}
  P_\mathrm{f}^i = (V^i - a^iP_\mathrm{c} - c^i)/b^i,
\end{equation}\end{linenomath}
where superscript indices range from $i=1,\ldots,4$ and correspond to each individual transducer FB1 to FB4. The estimated fluid pressure compares well to the one measured with the up- and downstream pressure transducers, with a maximum absolute deviation of around $0.5$~MPa over the whole $P_\mathrm{c}$ and $P_\mathrm{p}$ range (Figure \ref{fig:calibration}c, Table \ref{tab:coefs}). More specifically, at a given constant confining pressure, the precision of the transducer is of the order of $0.1$~MPa (Figure \ref{fig:calibration}c) over a $75$~MPa range in pore pressure, i.e., a relative precision of $0.13$\%. Therefore, if we restrict our experimentation to narrow ranges of fluid and confining pressure, we expect to be able to detect variations in $P_\mathrm{p}$ of the order of a few kPa, for sufficient amplification of the transducer outputs.

\section{Measurements of transport properties in intact and faulted sandstone}
\label{sec:ss}

\subsection{Deformation experiment in Darley Dale sandstone}

After the calibration procedure outlined above, the sample of Darley Dale sandstone was brought to $P_\mathrm{c}=100$~MPa and $P_\mathrm{p}=60$~MPa, and deformed at an axial strain rate of $10^{-5}$~s$^{-1}$ (Figure \ref{fig:deformation}). After an initial nonlinear adjustment phase, the sample behaved elastically and experienced compaction up to around 0.7\% axial strain. The pore volume change then deviated from a linear behaviour and showed increasing dilation up to failure at 1.4\% axial deformation. The sample experienced a peak stress at 183~MPa, and a rapid stress drop down to 106~MPa, which marks the formation of a macroscopic fault (Figure \ref{fig:deformation}c). No significant pore volume change was observed in association with the stress drop. The sample was then unloaded, pore pressure was dropped and confining pressure was decreased stepwise to 10~MPa. The sample was then brought again to $P_\mathrm{c}=100$~MPa and $P_\mathrm{p}=60$~MPa, and the sample was further deformed.

During the reloading phase, the sample also initially experienced compaction with increasing differential stress, followed by dilation and a stabilisation of the stress level at around 115~MPa. During further deformation, the sample showed continuous dilation at a rather modest rate (around +0.1\% porosity over 2.4\% axial deformation). The axial deformation rate was stepped up and down by a factor of 10 (marked as ``+'' and ``--'' signs in Figure \ref{fig:deformation}a), and only a mild rate-strengthening was observed. Increasing deformation rate did not produce significant deviation from the dilatant behaviour observed at $10^{-5}$~s$^{-1}$, whereas volume-neutral behaviour was observed at $10^{-6}$~s$^{-1}$.

Throughout all deformation, no significant pore pressure variation could be observed inside the sample, indicating a drained behaviour.

\subsection{Permeability heterogeneity}

At each pressure step during the first confining pressure cycle of the Darley Dale sandstone sample, fluid flow was imposed by venting the downstream end of the sample to the atmosphere and controlling the upstream end at a constant pore pressure of around 1.5~MPa. When steady-state was achieved, Darcy's law was used to estimate the average permeability of the sample $k_\mathrm{av}$. This procedure was repeated during a depressurisation cycle performed immediately after deformation and failure of the sample, and after further sliding along the newly formed fault by about $4$~mm.

\begin{figure}
  \centering
  \includegraphics{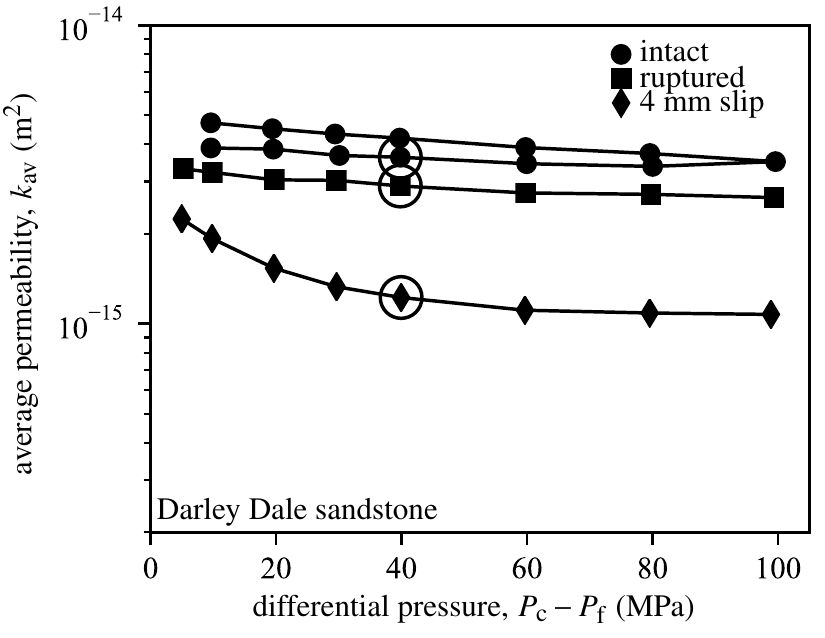}
  \caption{Average permeability measured by steady-state Darcy flow method in the intact Darley Dale sandstone sample (circles), immediately after shear rupture (squares) and after 4~mm slip across the newly formed fault (diamonds). The permeability was measured with both increasing and decreasing confining pressure steps in the intact sample, and only during decreasing confining pressure steps in the rupture and slipped sample. Pore pressure profiles during steady flow corresponding to circled data points are shown in Figure \ref{fig:ppprofiles}.}
  \label{fig:permeability}
\end{figure}

The average permeability of the intact sample is of around $5\times10^{-15}$~m$^2$ at $10$~MPa differential pressure, and decreases only slightly down to around $3.5\times10^{-15}$~m$^2$ when differential pressure is increased up to $100$~MPa. When pressure is decreased back to $10$~MPa, permeability remains essentially stable and returns to a value of around $4\times10^{-15}$~m$^2$ (Figure \ref{fig:permeability}). In the ruptured sample, permeability is only slightly lower, and is also mostly pressure-independent, at a value of around $3\times10^{-15}$~m$^2$. By contrast, the average sample permeability after 4~mm slip is significantly lower and exhibits pressure dependence: $k_\mathrm{av}$ is of around $2\times10^{-15}$~m$^2$ at $P_\mathrm{c}-P_\mathrm{f}=10$~MPa, and drops to $10^{-15}$~m$^2$ at $100$~MPa.

\begin{figure*}
  \centering
  \includegraphics{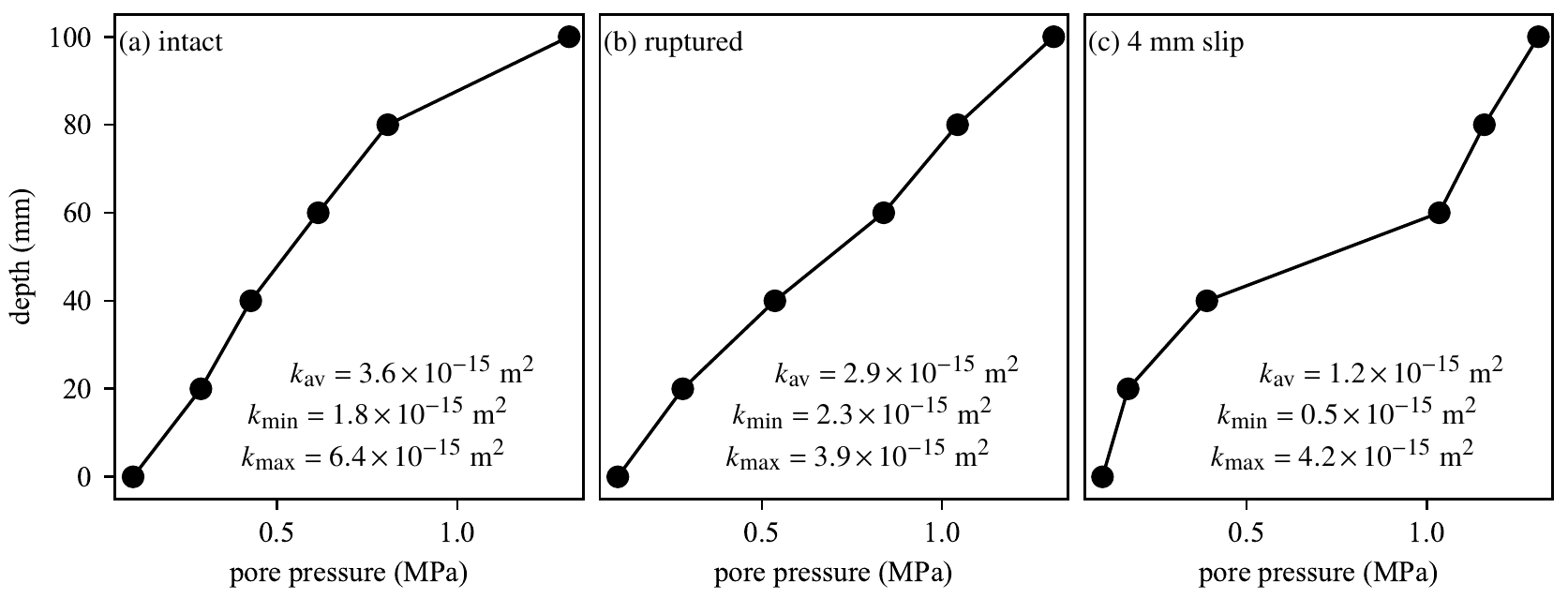}
  \caption{Pore pressure profiles during steady flow at $40$~MPa differential pressure (circled data points in Figure \ref{fig:permeability}) for (a) the intact sample, (b) the freshly ruptured sample, and (c) after 4~mm slip on the fault. The location of the transducers with respect to the fault is shown in Figure \ref{fig:deformation}c.}
  \label{fig:ppprofiles}
\end{figure*}

The pore pressure sensors can be used to examine the pore pressure profiles within the sample during steady flow. Figure \ref{fig:ppprofiles} shows representative examples of such profiles obtained at $40$~MPa differential pressure at the three different deformation stages of the sandstone sample. If the sample permeability was perfectly homogeneous, one should observe linear pore pressure profiles during steady flow. In the intact sample (Figure \ref{fig:ppprofiles}a), the pore pressure increases linearly from the downstream end up to the upper $80$~mm of the sample, and a slightly higher gradient is observed across the top $20$~mm. The permeability computed for each $20$~mm depth interval ranges from $1.8\times10^{-15}$~m$^2$ (in the upper part of the sample) to $6.4\times10^{-15}$~m$^2$ (in the lower part of the sample), showing some small preexisting heterogeneity. Immediately after rupture (prior to significant slip across the fault), the pore pressure profile is essentially linear across the whole sample, with permeability ranging from $2.3$ to $3.9\times10^{-15}$~m$^2$. The fault is located between the two middle transducers (the lower one of the pair being located almost on the fault but slightly below it, see Figure \ref{fig:deformation}c), but does not seem to have a significant impact on permeability at this stage. By contrast, after 4~mm slip (Figure \ref{fig:ppprofiles}c), the pore pressure profile essentially consists in two linear branches below and above the fault, and a steep gradient across the fault. The permeability across the fault  is significantly lower than in the lower and upper part of sample, at a minimum value of $0.5\times10^{-15}$~m$^2$.

\begin{figure}
  \centering
  \includegraphics{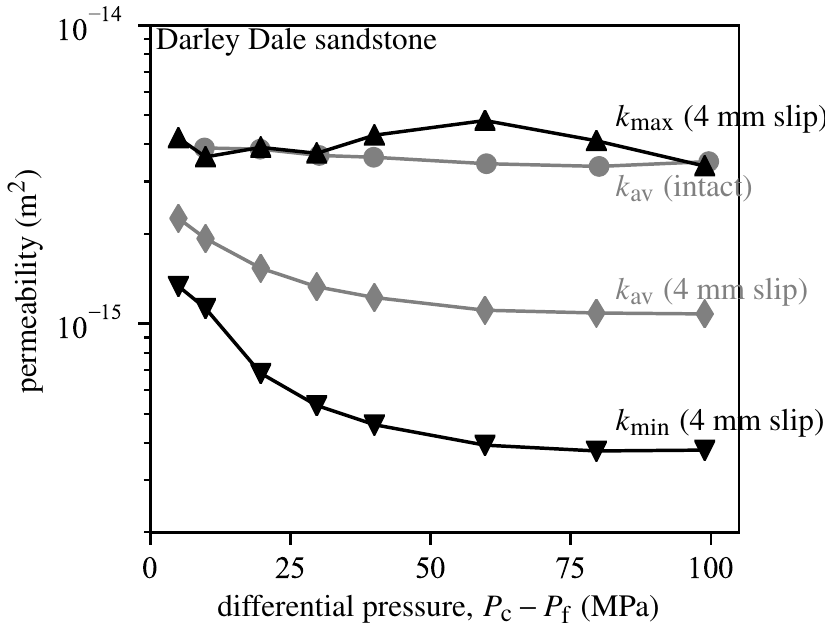}
  \caption{Evolution of local minimum and maximum permeability as a function of differential pressure in faulted Darley Dale sandstone after 4~mm slip, and comparison with average permeability in the intact sample.}
  \label{fig:kminPeff}
\end{figure}

Leveraging our ability to compute local permeability between pairs of transducers, we can now explore how the permeability heterogeneity evolves as a function of differential pressure in the faulted sample after 4~mm slip (Figure \ref{fig:kminPeff}). The maximum permeability $k_\mathrm{max}$ (measured off the fault) remains essentially pressure-independent and closely follows the average permeability measured in the intact material, which indicates that the fault walls retain the same properties as the intact rock. However, the minimum permeability $k_\mathrm{min}$ (measured across the fault) is not only much lower than that of the intact material, but also strongly pressure dependent, with values ranging from less than $0.4\times10^{-15}$~m$^2$ at $100$~MPa differential pressure up to around $10^{-15}$~m$^2$ at $10$~MPa.

The data indicate that the shear fault in Darley Dale sandstone becomes a permeability barrier after significant slip has been accumulated. Such a decrease is associated with the formation of fine-grained gouge within the fault, which increases the tortuosity of the flow paths \citep{zhu96,zhu97}. In addition, our novel instrumentation allows us to track local permeability heterogeneities, which reveals that the shear fault permeability is strongly pressure-dependent, whereas the intact rock is not. This observation can be explained by the increased compliance of the gouge material, which contains a network of thin cracks, in comparison with the relatively stiff pore network present in the intact material.

\section{Measurement of transport properties by the pore pressure step method}

In the previous Section we examined how local pressure measurements provide new constraints on the development of permeability heterogeneity in fractured rocks. The measurement of permeability was done using a steady flow method, and the internal pore pressure transducers were used to analyse the local pore pressure gradient.

In this Section, we use our internal pore pressure measurements to analyse the development of transient pore pressure fronts in a low permeability granite sample. The transient method used here is similar to the pulse-decay method \citep[e.g.,][]{brace68b,hsieh81,neuzil81,bourbie82}. We show that additional internal measurements are able to capture the full space-time evolution of the pore pressure field, consistent with the permeability structure of the sample.

\subsection{Model and principle}

Let us consider a sample of homogeneous porous, permeable material of length $L$ and cross sectional area $A$, saturated with a fluid of compressibility $\beta_\mathrm{f}$ at an initial homogeneous pore pressure equal $p_0$. The sample is connected to a downstream reservoir (at position $z=L$) of storage capacity $\beta_\mathrm{res}$, and to an upstream intensifier which imposes a pressure step $\Delta p$ at time $t=0^+$. The pore pressure evolution within the sample is given by the diffusion equation:
\begin{linenomath}\begin{equation}\label{eq:diff}
  \frac{\partial p}{\partial t} - \alpha \frac{\partial^2p}{\partial z^2} = 0,
\end{equation}\end{linenomath}
with boundary conditions
\begin{linenomath}\begin{equation}\label{eq:BCtop}
  p(0,t) = \Delta p + p_0
\end{equation}\end{linenomath}
and
\begin{linenomath}\begin{equation}\label{eq:BCbottom}
  \frac{\partial p}{\partial t} + \frac{k A}{\eta \beta_\mathrm{res}}\frac{\partial p}{\partial z} =0\quad\text{at } z=L,
\end{equation}\end{linenomath}
where we recall that $\alpha$ denotes the hydraulic diffusivity of the sample, expressed as a function of permeability $k$, fluid viscosity $\eta$ and storage capacity $\beta$ as $\alpha=k/(\eta/\beta)$. Equation \eqref{eq:diff} subject to boundary conditions \eqref{eq:BCtop} and \eqref{eq:BCbottom} has the following closed-form solution\citep[][Chap. 3, p. 128, problem (v)]{carslaw59}:
\begin{linenomath}\begin{equation}\label{eq:p}
  \frac{p(z,t) - p_0}{\Delta p} = 1 - \sum_{m=1}^\infty\frac{2(\varphi_m^2+\ell^2)\exp\big((-\varphi_m^2/\ell)\times(t/\theta)\big)\sin(\varphi_m z/L)}{\varphi_m(\varphi_m^2 + \ell^2 + \ell)},
\end{equation}\end{linenomath}
where $\theta=L\eta\beta_\mathrm{res}/(k A)$ is a characteristic diffusion time, $\ell = AL\beta/\beta_\mathrm{res}$ is the ratio of total sample storage capacity and the downstream storage capacity and the $\varphi_m$ are the roots of
\begin{linenomath}\begin{equation}
  \varphi\tan\varphi=\ell.
\end{equation}\end{linenomath}

The roots $\varphi_m$ can be ordered so that they each belong the interval $[(m-1)\pi, (m-1/2)\pi[$. At large time $t\gg\theta$, the closed-form solution \eqref{eq:p} is well approximated by
\begin{linenomath}\begin{equation} \label{eq:plarget}
  \frac{p(z,t)-p_0}{\Delta p} \approx 1 - \frac{2(\varphi_1^2+\ell^2)\exp\big[(-\varphi_1^2/\ell)\times(t/\theta)\big]\sin(\varphi_1 z/L)}{\varphi_1(\varphi_1^2 + \ell^2 + \ell)}.
\end{equation}\end{linenomath}
Furthermore, if $\ell\ll1$, then $\varphi_1\approx\sqrt{\ell}$, and \eqref{eq:plarget} is further approximated by
\begin{linenomath}\begin{equation}\label{eq:plarget_simple}
  \frac{p(z,t)-p_0}{\Delta p} \approx 1 - \frac{z}{L}\exp(-t/\theta),\qquad t/\theta\gg1\quad\text{and}\quad \ell\ll1.
\end{equation}\end{linenomath}
This approximation highlights the fact that if the sample's total storage capacity is much smaller than that of the downstream reservoir, a likely situation in laboratory experiments, then the pore pressure evolution at large time depends only on the position $z/L$ and the dimensionless time $t/\theta$, which is independent from $\beta$. Therefore, pore pressure records do not contain information about the sample's storage capacity at sufficiently long timescales.

By contrast, at small time, the solution \eqref{eq:p} is well approximated by (see Appendix \ref{ax:1})
\begin{linenomath}\begin{equation}\label{eq:psmallt}
  \frac{p(z,t)-p_0}{\Delta p} \approx \erfc\left(\frac{z/L}{2\sqrt{t/\tau}}\right) + \erfc\left(\frac{2-z/L}{2\sqrt{t/\tau}}\right) - 2\exp\big(\ell(2-z/L)+\ell^2t/\tau\big)\erfc\left(\frac{2-z/L}{2\sqrt{t/\tau}}+\ell\sqrt{t/\tau}\right),
\end{equation}\end{linenomath}
where $\tau=L^2/\alpha$ is the characteristic diffusion time in the sample. Further considering the case $\ell\ll1$, the small time approximation simplifies to
\begin{linenomath}\begin{equation} \label{eq:psmallt_simple}
  \frac{p(z,t)-p_0}{\Delta p} \approx \erfc\left(\frac{z/L}{2\sqrt{t/\tau}}\right)  - \erfc\left(\frac{2-z/L}{2\sqrt{t/\tau}}\right),\qquad t/\tau\ll1\quad\text{and}\quad \ell\ll1.
\end{equation}\end{linenomath}
The small time approximation clearly shows a dependence in both permeability and storage capacity through the diffusion time $\tau$, and the pore pressure records at short timescales are therefore expected to contain information on both parameters. One peculiarity of the small time solution \eqref{eq:psmallt_simple} is that the pore pressure change at $z=L$, i.e., at the downstream end of the sample, remains zero to first order, while this is not the case at $z<L$. Thus, we expect that recording pore pressure at different $z$ should bring independent constraints on the diffusivity $\alpha$.

\begin{figure*}
  \centering
  \includegraphics{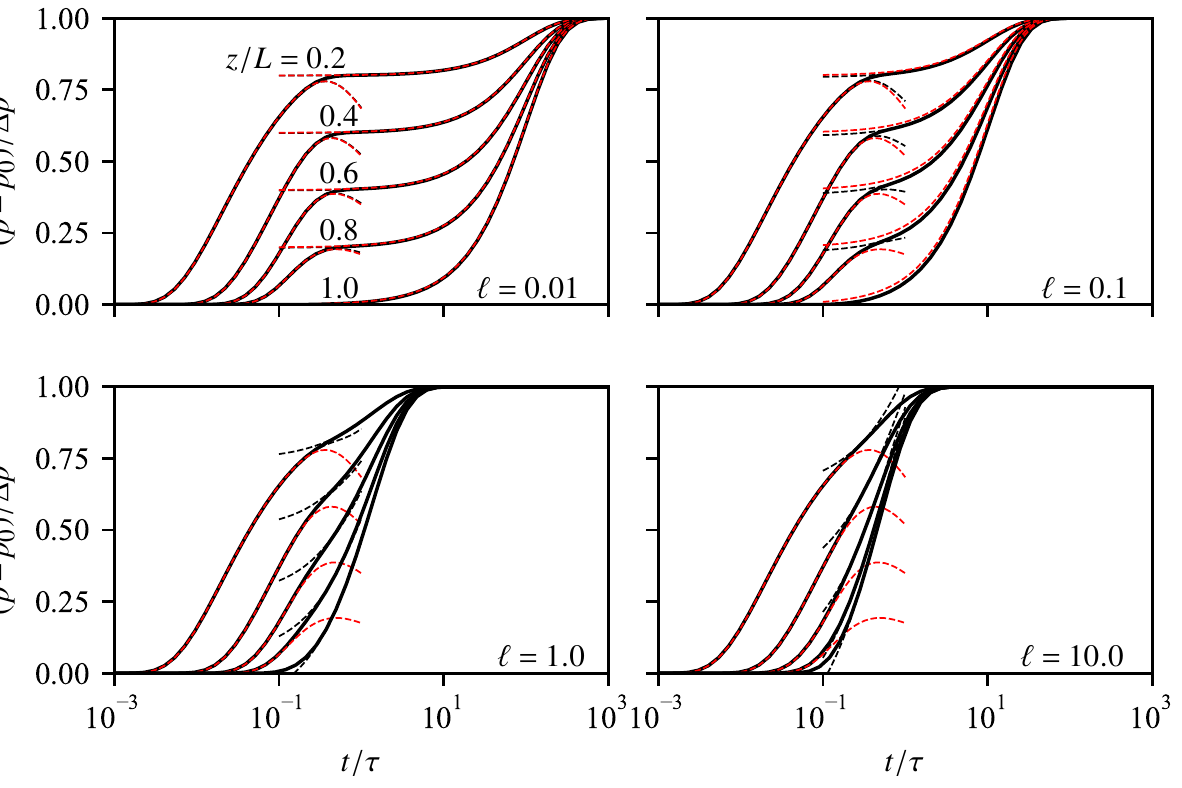}
   \caption{Pore pressure evolution at different positions $z/L$ as a function of time (normalised by the diffusion time $\tau=L^2/\alpha$) for a range of normalised storage $\ell = AL\beta/\beta_\mathrm{res}$. Black lines correspond to the full closed-form solution \eqref{eq:p}, dashed black lines correspond to the small and large time solutions \eqref{eq:psmallt} and \eqref{eq:plarget}, and red dashed lines correspond to the simpler approximations of the asymptotic cases \eqref{eq:psmallt_simple} and \eqref{eq:plarget_simple}.}
  \label{fig:pstep}
\end{figure*}

Figure \ref{fig:pstep} shows the general evolution of pore pressure at different positions inside the sample, and illustrates that the simple approximations for small and large time (Equations \ref{eq:psmallt_simple} and \ref{eq:plarget_simple}) are remarkably effective in cases when $\ell\lesssim0.1$.

From the full solution \eqref{eq:p}, the cumulative fluid volume change at the upstream end of the sample ($z/L=0$) is obtained by direct calculation as
\begin{linenomath}\begin{equation}\label{eq:flow}
  \Delta v(t) = \Delta p\beta_\mathrm{res}\sum_{m=1}^\infty\frac{2\ell(\varphi_m^2+\ell^2)\big[1-\exp\big((-\varphi_m^2/\ell)\times(t/\theta)\big)\big]}{\varphi_m^2(\varphi_m^2 + \ell^2 + \ell)}.
\end{equation}\end{linenomath}
By similar arguments to those explained above for the key controls on pore pressure evolution at small and large time, we also expect that measurements of flow volume would only contain information about $\theta$ at large time and for $\ell\ll1$, and are dominated by diffusivity (through the timescale $\tau$) at small time. However, measuring flow rate does not yield only redundant information. Indeed, as $t\rightarrow\infty$, the fluid volume change tends to $\Delta v \rightarrow \Delta p \beta_\mathrm{res}(1+\ell)$, which is simply equivalent to $\Delta v \rightarrow \Delta p (\beta_\mathrm{res} + AL\beta)$. Thus, if $\ell$ is not much smaller than $1$, fluid volume change can potentially lead to a direct estimation of storage capacity. We will illustrate below that combining both pressure and flow measurements allows to determine both permeability and storage capacity.

\begin{figure}
  \centering
  \includegraphics{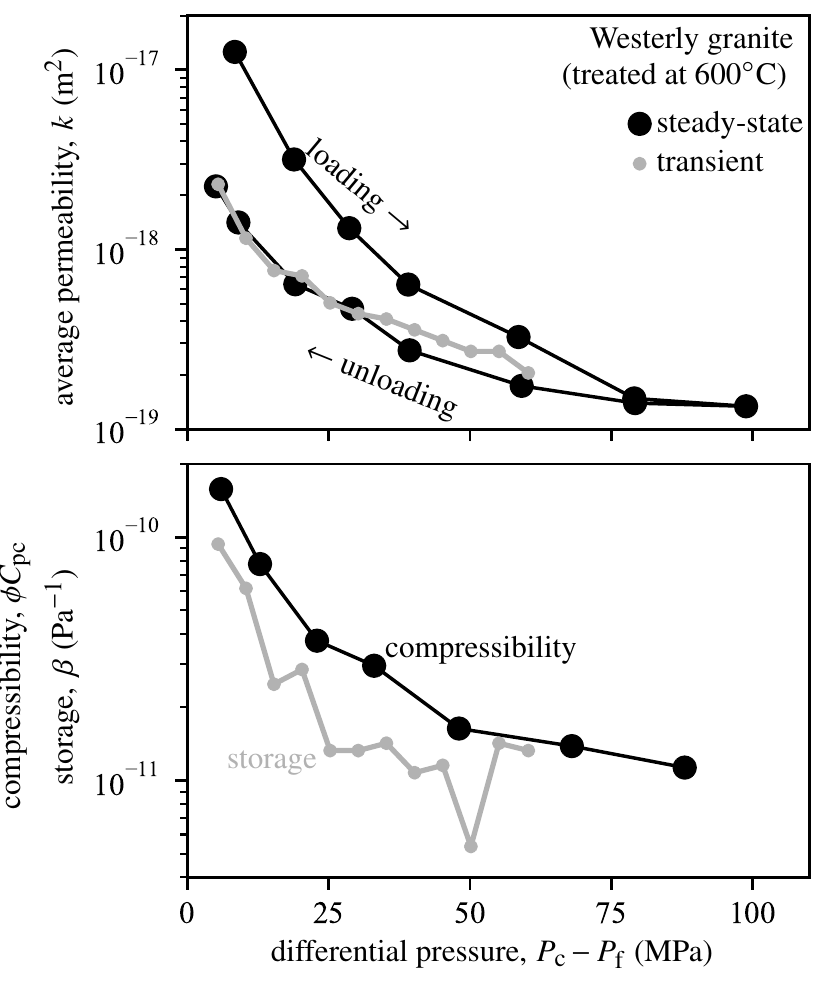}
  \caption{Permeability and compressibility evolution in Westerly granite thermally cracked at $600^\circ$C. Black circles are permeability values measured using the constant flow rate method, and compressibility values computed from the change in total pore volume in response to changes in confining pressure. Grey dots are permeability and storage capacity values measured using the transient step method.}
  \label{fig:wgproperties}
\end{figure}

\subsection{Experimental method and initial characterisation}

The sample of thermally cracked Westerly granite was tested under hydrostatic conditions at $P_\mathrm{c}$ from 5 to 100~MPa. During the first loading and unloading stage, the average sample permeability was measured using the constant flow rate method, imposing an upstream pore pressure of 2.2 to 3.1~MPa and venting the downstream end of the sample. During unloading, the total pore volume change associated with the decrease in confining pressure was recorded, and the compressibility $C_\mathrm{bp} = \phi C_\mathrm{pc} = (1/V_\mathrm{bulk})\partial V_\mathrm{pore}/\partial P_\mathrm{c}$, where $\phi$ is the sample's porosity, $V_\mathrm{bulk}$ its bulk volume and $V_\mathrm{pore}=\phi V_\mathrm{bulk}$ \citep[][Chap. 7]{jaeger07} was computed. Permeability is initially of the order of $k=10^{-17}$~m$^2$ at 10~MPa differential pressure, and decreases to around $10^{-19}$~m$^2$ at 100~MPa. During unloading, permeability increases but does not recover entirely, reaching only around $2\times10^{-18}$~m$^2$ when differential pressure is reduced to 5~MPa. Concomitantly, the compressibility $\phi C_\mathrm{pc}$ increases from around $10^{-11}$~Pa$^{-1}$ up to $1.5\times10^{-10}$~Pa$^{-1}$. 

\begin{figure}
  \centering
  \includegraphics{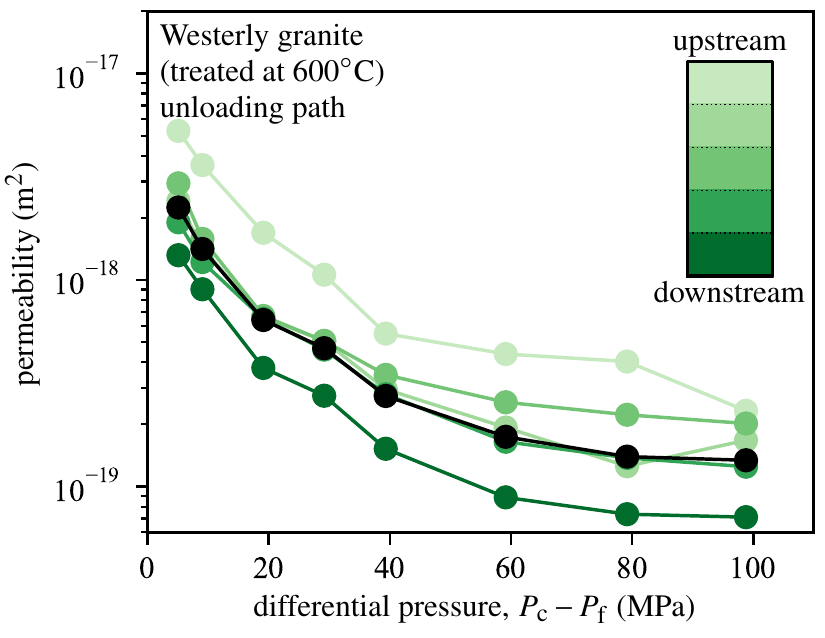}
  \caption{Permeability evolution in thermally cracked Westerly granite as a function of differential pressure during hydrostatic unloading. Each color correspond to the average permeability in the layer shown in the inset. Black line is the sample-averaged permeability (shown in Figure \ref{fig:wgproperties}a).}
  \label{fig:wgperm}
\end{figure}

During constant flow rate tests, we used the stead-state pore pressure profiles recorded by the four internal and up- and downstream transducers to extract the local permeability value in five 20~mm-width layers along the sample height. The sample is not perfectly homogeneous (Figure \ref{fig:wgperm}). At any given pressure, the permeability in the upper part of the sample (top 20~mm layer) is larger by a factor 2.0 to 2.9 compared to the sample-averaged permeability, while the bottom part of the sample (bottom 20~mm layer) is smaller by a factor 1.5 to 1.9. The central part of the sample has a permeability close to the average. The permeability has the same pressure sensitivity in all layers.

After the first confining pressure unloading stage, the confining pressure was increased again to $P_\mathrm{c}=75$~MPa, and the pore pressure was equilibrated throughout the sample at $P_\mathrm{f}=5$~MPa. Then, the downstream end of the sample was isolated, the high pressure pipe volume forming a closed reservoir of calibrated storage capacity $\beta_\mathrm{res} = 6\times10^{-15}$~m$^3$Pa$^{-1}$. The pore pressure at the upstream end was then increased by a $5$~MPa increment (ramp duration of a few seconds), and we recorded the time evolution of pore pressure within the sample using four internal pore pressure transducers, and in the downstream reservoir using a conventional pressure transducer connected to the pipe volume. The fluid volume change was also recorded at the upstream end. The intensifier volume initially changed in proportion to the pressure change (due to upstream storage), which could be measured as soon as the target upstream was reached. The fluid volume change was corrected from this initial change, assuming negligible flow into the sample during the pressure ramp (only a few seconds). When the pore pressure fully re equilibrated, a new step was imposed upstream. The procedure was repeated up to $P_\mathrm{f}=70$~MPa.

\subsection{Experimental results using transient step}

\begin{figure*}
  \centering
  \includegraphics{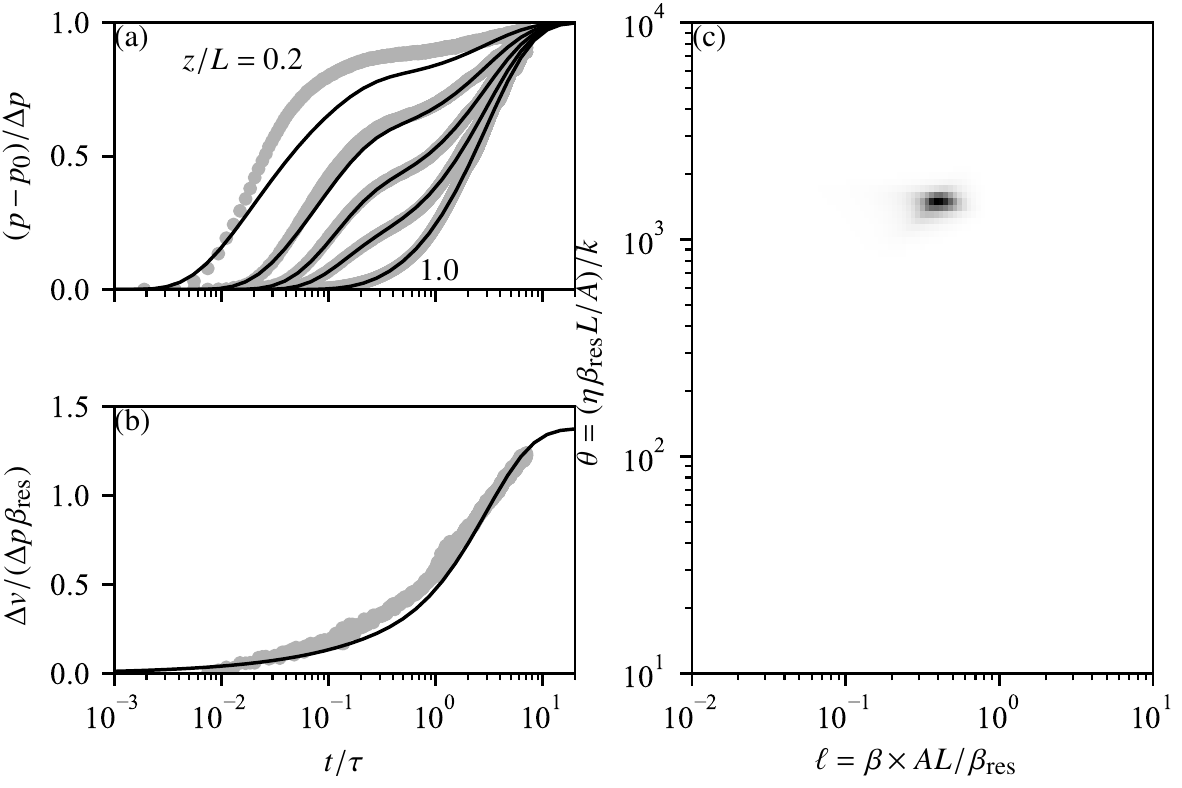}
  \caption{Normalised pore pressure at positions $z/L=0.2,\ldots,1.0$ (a) and intensifier volume (b) as a function of time during a step test conducted on thermally cracked Westerly granite at $P_\mathrm{c}=75$~MPa, initial pore pressure $p_0=15$~MPa and upstream step amplitude $\Delta p = 5$~MPa. Grey points are data, and solid black lines are best fit to Equation \eqref{eq:p}. Panel (c) shows the probability density in $(\theta,\ell)$ space computed using a full parameter exploration and a least-absolute value criterion for the misfit between data and model predictions.}
  \label{fig:example_pstep}
\end{figure*}

A representative example of pore pressure and fluid volume time series recorded during a given upstream pore pressure step is shown in Figure \ref{fig:example_pstep} (panels a and b, grey dots). The general trend of the data is in qualitative agreement with the features highlighted in Figure \ref{fig:pstep}: the pore pressure measured at the downstream reservoir ($z/L=1.0$) follows a single increasing trend, while the pore pressure measured at different positions along the sample height exhibit an initial rapid rise followed by a plateau and a subsequent regular increase as pore pressure becomes uniform within the sample.

The whole dataset was used to find best-fitting values of parameters $\ell$ and $\theta$. The pore pressure and pore volume data were first down-sampled to a uniformly spaced time array with 200 steps. A full grid search was conducted in the $(\theta,\ell)$ space, and for each parameter combination the misfit between the data and model prediction was computed using the sum of the absolute differences between data and model points. All misfits were weighted by a unique error on the data chosen to be $100$ for pore pressure and $10$ for fluid volume (in normalised pressure and pore volume units). These very large artificial errors account for imperfection in the theoretical model and other imposed parameters such as sensor positions \citep[][p. 23]{tarantola05}. The weighted misfits are then used to produce a map of probability density (assuming double laplacian errors on the data) in the $(\ell,\theta)$ space, shown in Figure \ref{fig:example_pstep}c. The peak probability is unique and corresponds to a relatively narrow range of both $\theta$ and $\ell$, which implies a unique pair of permeability and sample storage capacity. The theoretical predictions using the best fit parameters, shown as solid black lines in panels a and b, are in very good agreement with the data. One significant deviation, observed systematically across our dataset, is that the pore pressure measured in the uppermost transducer (position $z/L=0.2$) is always larger than predicted. This discrepancy is consistent with the larger permeability measured in the upper part of the sample using the steady flow method (Figure \ref{fig:wgperm}).

The data for all pore pressure steps were inverted to obtain best-fitting values of permeability (through parameter $\theta$) and storage capacity (through parameter $\ell$) as a function of differential pressure $P_\mathrm{c}-P_\mathrm{f}$, shown in Figure \ref{fig:wgproperties} (grey dots). The inferred average permeability is in very good quantitative agreement with that measured using steady-state flow along the unloading branch. Unlike the steady-flow method, the transient method does not allow for a simple characterisation of permeability heterogeneity, and the permeability estimate is close to the sample average despite the existence of local variations. The larger permeability in the top layer is only seen qualitatively by direct comparison of the pore pressure records with theoretical predictions (Figure \ref{fig:example_pstep}a). In addition, the lower permeability in the bottom layer is not detectable in the transient records: Since the pore pressure step is imposed at the upstream end, the downstream pressure record reflects the average sample permeability rather than the local one.

The storage capacity is of similar order of magnitude as the measured compressibility $\phi C_\mathrm{pc}$ and follows a similar trend, but appears systematically lower. This is not consistent with the definition of storage capacity, which is expressed as
\begin{linenomath}\begin{equation}\label{eq:storage}
  \beta = \phi(\beta_\mathrm{f} + C_\mathrm{pc} - C_\mathrm{m}),
\end{equation}\end{linenomath}
where $\beta_\mathrm{f}$ is the fluid compressibility and $C_\mathrm{m}$ is the solid matrix compressibility. Neglecting the contribution of the latter, we should expect $\beta$ to be always larger than $\phi C_\mathrm{pc}$, by an amount given by $\phi \beta_\mathrm{f}$, of the order of $5\times10^{-12}$~Pa$^{-1}$ (computed for $\phi=0.01$). There are several possible reasons for the discrepancy between our data and the theory. Firstly, the measurement of compressibility is performed using relatively large steps of confining pressure (5 to 10~MPa), so that there might be a significant contribution of inelastic porosity change, which would lead to overestimating the elastic pore space compressibility. In addition, there might be nonzero contributions of intra-vessel tubing volume change during confining pressure steps. Second, the storage capacity expressed in Equation \eqref{eq:storage} is obtained under constant stress boundary conditions \citep[][Chap. 7]{jaeger07}; however, during pore pressure steps, friction at the contact between sample ends and steel end-caps limits lateral expansion of the sample and the top and bottom boundaries are more likely under constant strain conditions. The storage capacity under constant strain conditions is always smaller than that under constant stress conditions \citep[][Section 3.3]{wang00}, which could explain the lower values inferred from our transient step measurements.


Overall, the results from transient tests illustrate the ability of our new internal pore pressure transducers to track the development of pore pressure variations inside rock samples under high pressure conditions. The internal measurements are consistent with theoretical predictions and can be combined in an inverse problem to recover accurately the average permeability of the samples, together with an estimate of their storage capacity.

\section{Summary and conclusions}

We presented the design of a new low volume pore pressure transducer that can be mounted in direct contact with rock samples inside a triaxial apparatus. The transducer effectively measures the difference between confining and pore pressure, and its response can be easily calibrated by imposing known combinations of confining and pore fluid pressure. All our manufactured transducers have a very good linear response over a $0$ to $80$~MPa pressure range, with absolute errors of the order of $0.25$~MPa. The low ``dead'' volume of the transducers allows for a fast response time in most porous rocks; in very tight materials with low permeability and low storage, the transducer response time can be as high as a few seconds (Table \ref{tab:filterrocks}).

We demonstrated the use of our transducers in a triaxial deformation experiment conducted on Darley Dale sandstone. Using an array of four transducers regularly spaced across the sample, in addition to more conventional upstream and downstream pressure measurements, we showed that the sample was initially relatively homogeneous and pressure-insensitive. We deformed the sample at $40$~MPa differential pressure, which lead to shear failure and slip on a fault. During steady flow across the fault, the pore pressure measurements showed a strong local decrease in permeability across the fault. The  permeability of the fault zone material exhibited a significant pressure dependency, whereas that of the wall rock remained pressure independent and equal to the intact value.

In addition, we tested our transducers on a thermally cracked Westerly granite sample subjected to a sequence of pore pressure steps. Each step was imposed only at the upstream end of the sample, while the downstream end remained closed to a fixed volume reservoir. The internal pore pressure measurements were shown to be in very close agreement with the theoretical solution, illustrating the transient propagation of the pore pressure front. Our analysis confirms that of \citet{nicolas20}, who measured pore pressure fronts in cracked andesite using fibre optic pressure sensors. The complete dataset of internal and downstream pore pressure measurements, together with upstream fluid volume change, was used in an inverse problem to obtain joint estimates of permeability and storage capacity. The benefit of internal pore pressure measurements is that their behaviour at small time is dominated by hydraulic diffusivity, whereas the downstream pore pressure is only sensitive to permeability (in tight rocks). Our inverted values of permeability were found to be consistent with permeability values obtained from constant flow rate method. The inferred storage capacity appeared to be slightly underestimated when compared to an independent measure of pore space compressibility, but remained of a correct order of magnitude and trend.

The transducers presented here are remarkably cheap, robust, and easy to use. Their footprint is small, similar to that of typical acoustic emission or ultrasonic transducers, and a possibly large array of transducers can be positioned around a given sample. This opens new avenues to better understand local variations in fluid pressure and heterogeneities during deformation and fluid flow in rocks. Several applications in different areas of rock physics are currently being studied: the negligible transducer volume and fast response time allows for nearly perfectly undrained conditions to be met in experiments, which allows for measurements of poroelastic properties; fault zone hydraulic properties can also now be measured in situ during fault slip, which has great potential to unravel the complexity of hydromechanical couplings during the seismic cycle \citep{proctor20,brantut20}; finally, systematic measurements of permeability heterogeneity due to strain localisation in porous materials, as illustrated here in Section \ref{sec:ss}, can now be undertaken.

\begin{acknowledgments}
  Neil Hughes contributed to the design and machining of the prototypes. Discussions with Philip Meredith and Tom Mitchell helped shape this project. Comments from two reviewers helped clarify the manuscript. We acknowledge funding from the UK Natural Environment Research Council (grants NE/K009656/1, NE/M016471/1 and NE/S000852/1) and from the European Research Council under the European Union's Horizon 2020 research and innovation programme (project RockDEaF, grant agreement \#804685). Experimental data can be found at the NGDC repository of the British Geological Survey (ID 136076). 
\end{acknowledgments}

\appendix

\section{Pressure disturbance due to transducer in response to step change}
\label{ax:0}

\subsection{Numerical solution}

Equation \eqref{eq:dp_2d} together with boundary condition \eqref{eq:BCdim} over the region $|r|<a$ and $\partial p/\partial z =0$ in the region $|r|>a$ was solved numerically by using a fully implicit finite difference method. Time was normalised by $a^2/\alpha$ and space by $a$. Space was discretized into a nonuniform grid, with $71$ nodes in the $r$ coordinate including 21 equally spaced nodes in $|r|/a<1$ and 50 logarithmically spaced nodes from $|r|/a=1$ to $|r|/a=30$; 71 logarithmically spaced nodes were used along the $z$ coordinate. Time was discretised in uniform steps of at most $1.25\times10^{-3}$.

\subsection{One dimensional solution}

Neglecting flow in the radial direction, and using the normalisation $t\leftarrow t/\tau_\mathrm{trans}$, $z\leftarrow z/\sqrt{\alpha\tau_\mathrm{trans}}$ and $p\leftarrow (p-p_0)/\Delta p$, Equation \eqref{eq:dp_2d} is rewritten as
\begin{linenomath}\begin{equation}\label{eq:diffnorm1}
  \frac{\partial p}{\partial t} - \frac{\partial^2p}{\partial z^2} = 0
\end{equation}\end{linenomath}
and the boundary condition is expressed as
\begin{linenomath}\begin{equation}\label{eq:BCnorm1}
  \frac{\partial p}{\partial t} + \frac{\partial p}{\partial z} =0\quad\text{at } z=1.
\end{equation}\end{linenomath}
Denoting $\tilde{p}(z,s)$ the Laplace transform of $p(z,t)$, 
\begin{linenomath}\begin{equation}
  \tilde{p}(z,s) = \int_0^\infty p(z,t)e^{-st}dt,
\end{equation}\end{linenomath}
the Laplace transforms of Equations \eqref{eq:diffnorm1} and \eqref{eq:BCnorm1} are given by
\begin{linenomath}\begin{align}
s\tilde{p} - 1 - \frac{d^2 \tilde{p}}{dz^2} &= 0,\\
s\tilde{p} + \ell\frac{d\tilde{p}}{dz} &= 0 \text{ at } z=1,
\end{align}
\end{linenomath}
respectively. The solution for $\tilde{p}$ in the Laplace domain is given by conventional methods \citep[e.g.,][]{carslaw59,hsieh81} as
\begin{linenomath}\begin{equation}
  \tilde{p}(z,s) = \frac{1}{s} - \frac{e^{-qz}}{s+q},
\end{equation}\end{linenomath}
where $q=\sqrt{s}$. A direct inversion of the Laplace transform leads to \citep[][Chap. 13]{carslaw59}:
\begin{linenomath}\begin{equation}
  p(z,t) =  1 - e^{z+t}\mathrm{erfc}\big(\frac{z}{2\sqrt{t}}+\sqrt{t}\big).
\end{equation}\end{linenomath}
The result at $z=0$ is given in dimensional form as:
\begin{linenomath}\begin{equation}
  \frac{p(0,t)-p_0}{\Delta p} =  1 - e^{t/\tau_\mathrm{trans}}\mathrm{erfc}\big(\sqrt{t/\tau_\mathrm{trans}}\big),
\end{equation}\end{linenomath}
and is plotted in Figure \ref{fig:filter} as dashed red lines.

\subsection{Membrane approximation}

The evolution of fluid pressure in the sensor cavity is given by the boundary condition \eqref{eq:BCdim}. The pore pressure far from the transducer is expected to be unaffected by its dead volume. By analogy with heat transfer problems, where the heat flux is assumed to be directly proportional to the temperature difference between two materials, we can approximate the pressure gradient by the ratio
\begin{linenomath}\begin{equation}
  \frac{\partial p}{\partial z} \approx \frac{p - p_0-\Delta p}{a},
\end{equation}\end{linenomath}
where the characteristic distance over which the gradient is established is assumed to be equal to the sensor radius $a$. This approximation is also called ``membrane diffusion'', and is typically used when fluid diffusion occurs across a thin zone with low diffusivity. Equation \eqref{eq:BCdim} can then be solved directly:
\begin{linenomath}\begin{equation}
  \frac{p(t) - p_0}{\Delta p} = 1 - \exp\left(-\frac{t}{a\beta_\mathrm{trans}\eta/(Ak)}\right),
\end{equation}\end{linenomath}
where we note that
\begin{linenomath}\begin{equation}
  a\beta_\mathrm{trans}\eta/(Ak) = \sqrt{\tau\tau_\mathrm{trans}}
\end{equation}\end{linenomath}
where $\tau = a^2/\alpha$ (see solution in Equation \ref{eq:p_membrane}).

The associated behaviour is remarkably similar to that obtained with fully numerical 2D solutions at large time (Figure \ref{fig:filter}), and gives a very simple estimate of the sensor response time as a function of sensor and rock parameters.

\section{Small time approximation of the pore pressure step problem}
\label{ax:1}

Using the normalisation $t\leftarrow t/\tau$, $z\leftarrow z/L$ and $p\leftarrow (p-p_0)/\Delta p$, Equation \eqref{eq:diff} is rewritten as
\begin{linenomath}\begin{equation}\label{eq:diffnorm}
  \frac{\partial p}{\partial t} - \frac{\partial^2p}{\partial z^2} = 0
\end{equation}\end{linenomath}  
and the boundary conditions are expressed as
\begin{linenomath}\begin{equation}\label{eq:BCnorm}
  p(0,t) = 1,\qquad \frac{\partial p}{\partial t} + \ell\frac{\partial p}{\partial z} =0\quad\text{at } z=1.
\end{equation}\end{linenomath}
Denoting $\tilde{p}(z,s)$ the Laplace transform of $p(z,t)$, 
\begin{linenomath}\begin{equation}
  \tilde{p}(z,s) = \int_0^\infty p(z,t)e^{-st}dt,
\end{equation}\end{linenomath}
the Laplace transforms of Equations \eqref{eq:diffnorm} and \eqref{eq:BCnorm} are given by
\begin{linenomath}\begin{align}
s\tilde{p} - \frac{d^2 \tilde{p}}{dz^2} &= 0,\\
\tilde{p}(0,s) &= 1/s,\\
s\tilde{p} + \ell\frac{d\tilde{p}}{dz} &= 0 \text{ at } z=1,
\end{align}
\end{linenomath}
respectively. The solution for $\tilde{p}$ in the Laplace domain is given by conventional methods \citep[e.g.,][]{carslaw59,hsieh81} as
\begin{linenomath}\begin{equation}
  \tilde{p}(z,s) = \frac{1}{s}\frac{(\ell q -s)e^{q(z-1)} + (\ell q +s)e^{-q(z-1)}}{(\ell q +s)e^{q} + (\ell q -s)e^{-q}},
\end{equation}\end{linenomath}
where $q=\sqrt{s}$. Rewriting the solution in Laplace domain as
\begin{linenomath}\begin{equation}
  \tilde{p}(z,s) = \frac{1}{s}\left[\frac{\ell q -s}{\ell q +s}e^{q(z-2)} + e^{-q(z-2)}\right]\times\frac{1}{1 + (\ell q -s)/(\ell q +s)e^{-2q}}
\end{equation}\end{linenomath}
and expanding the last term on the right hand side for large $q$, retaining the first three dominant terms, we obtain
\begin{linenomath}\begin{equation}
  \tilde{p}(z,s) \approx \frac{e^{-qy}}{s} - \frac{e^{q(z-2)}}{q(\ell+q)} + \ell\frac{e^{q(z-2)}}{s(\ell+q)}.
\end{equation}\end{linenomath}
This can be transformed into time domain using common Laplace transform tables \citep[][Appendix V]{carslaw59}, which yields the small time approximation \eqref{eq:psmallt}.


\balance

\end{document}